\documentclass[12pt]{elsarticle}

\RequirePackage[OT1]{fontenc}
\RequirePackage{amsthm,amsmath,amsfonts}
\RequirePackage{natbib}
\RequirePackage[colorlinks,citecolor=blue,urlcolor=blue]{hyperref}
\usepackage{tikz}
\usepackage{graphicx}
\usepackage[caption=false]{subfig}
\newtheorem{prop}{Proposition}

\newcommand{\blind}{0}

\usetikzlibrary{arrows,chains,matrix,positioning,scopes,fit}
\makeatletter
\tikzset{join/.code=\tikzset{after node path={%
			\ifx\tikzchainprevious\pgfutil@empty\else(\tikzchainprevious)%
			edge[every join]#1(\tikzchaincurrent)\fi}}}
\makeatother
\tikzset{>=stealth',every on chain/.append style={join},
	every join/.style={->}}
\tikzstyle{labeled}=[execute at begin node=$\scriptstyle,
execute at end node=$]
\usepackage{amssymb}

\begin{document}
	\begin{frontmatter}
	
	\def\spacingset#1{\renewcommand{\baselinestretch}%
		{#1}\small\normalsize} \spacingset{1}
	
\if0\blind
{
	\title{\bf An Extended Laplace Approximation Method for Bayesian Inference of Self-Exciting  Spatial-Temporal Models of Count Data}
	\author{Nicholas J. Clark\thanks{
			Supported in part through a fellowship from the \textit{The Omar N. Bradley Foundation}}\hspace{.2cm}\\
		Department of Statistics, Iowa State University\\
		and \\
		Philip M. Dixon \\
		Department of Statistics, Iowa State University}

} \fi

\if1\blind
{
	\bigskip
	\bigskip
	\bigskip
	\begin{center}
		{\LARGE\bf An Extended Laplace Approximation Method for Inference on Self-Exciting Poisson Spatial-Temporal Models}
	\end{center}
	\medskip
} \fi

\bigskip
\begin{abstract}

Self-Exciting models are statistical models of count data where the probability of an event occurring is influenced by the history of the process.  In particular, self-exciting spatio-temporal models allow for spatial dependence as well as temporal self-excitation.  For large spatial or temporal regions, however, the model leads to an intractable likelihood.  An increasingly common method for dealing with large spatio-temporal models is by using Laplace approximations (LA).  This method is convenient as it can easily be applied and is quickly implemented.  However, as we will demonstrate in this manuscript, when applied to self-exciting Poisson spatial-temporal models, Laplace Approximations result in a significant bias in estimating some parameters.  Due to this bias, we propose using up to sixth-order corrections to the LA for fitting these models.  We will demonstrate how to do this in a Bayesian setting for Self-Exciting Spatio-Temporal models.  We will further show there is a limited parameter space where the extended LA method still has bias.  In these uncommon instances we will demonstrate how amore computationally intensive fully Bayesian approach using the Stan software program is possible in those rare instances.  The performance of the extended LA method is illustrated with both simulation and real-world data.
\end{abstract}


\end{frontmatter}

\section{Introduction} 

Intractable likelihood functions arise in a multitude of settings in statistics, especially in modeling spatio-temporal data.  For spatial or spatio-temporal models it is oftentimes easier to specify the probability of an event occurring at a given location conditional on the occurrence or non-occurrence at neighboring events.  In this instance, it is easy to write down the conditional density, but the joint density may not have a closed form expression, or, if it does, the likelihood cannot be evaluated.  

 For example, in a spatial process observed on a fixed lattice we may have, writing $s_i \in \{s_1,s_2,...,s_{n_d}\}$ as fixed locations in $\mathbb{R}^2$, $Z(s_i)\sim \mbox{Pois}(\lambda(s_i))$ as observed counts at a given location.  We may further have $\boldsymbol{\lambda} \sim \mbox{Log Gau} (\boldsymbol{\alpha}, \Sigma(\theta))$ where $\boldsymbol{\lambda}$ is the vector of all Poisson expectations at each location and $\mbox{Log Gau}$ is the standard multivariate log-Gaussian distribution.Spatial structure may be placed on $\Sigma(\theta)$ by, for example, letting $\Sigma(\theta) = (I_{n_d,n_d}-C)^{-1}M$ where $I$ is the identity matrix, $C$ is a matrix with entries   $\zeta$ at location $i,j$ if spatial locations $s_i$ and $s_j$ are spatial neighbors, and $M$ is a diagonal matrix with diagonal entries, $\tau^2$.  This model is oftentimes called the Poisson-CAR model and is described in detail in Section 4.2 of \cite{cressie2015statistics}.  The log-likelihood for the spatial parameters is proportional to the intractable integral
 \begin{equation}
 l_{n_d}(\theta) \propto -\frac{1}{2}\log \det \left(\Sigma (\theta)\right) + \log \int_{\mathbb{R}^{n_d}} \exp \left(\sum_{s_i=1}^{s_i={n_d}} Z(s_i)Y(s_i)-\exp(Y(s_i)) - \frac{1}{2} \boldsymbol{Y}^T \Sigma^{-1}(\theta) \boldsymbol{Y} \right) \text{d} \boldsymbol{Y} \label{eq:loglik},
 \end{equation}
 where $\theta$ is the set of all spatial parameters.

However, while the integral in \eqref{eq:loglik} is intractable, it is of the form $I_n = \int_{\mathbb{R}^n} \exp(-h_{d}\left(Y\right)) \text{d}y$ allowing for Laplace approximations to be used to conduct inference.  In both spatial and spatio-temporal modeling, using Laplace approximations to conduct inference on the spatial or spatio-temporal diffusion parameters has dramatically increased since the advent of the Integrated Nested Laplace Approximation, or INLA, package from \cite{rue2009approximate}.  \cite{rue2017bayesian} provides many examples of INLA being used in literature.

Though the Laplace approximation technique is extremely fast compared to Markov Chain Monte Carlo (MCMC) techniques and it provides consistent estimates for parameters, it only does so asymptotically where the asymptotic error rate decreases as a function of pseudo independent observations.  By pseudo independent we mean observations that are separated sufficiently far in either spatial or temporal distance as to have minimal influence on one another.  For example, in \cite{cressie1992statistics} on page 15 it is shown how a simple spatial-only model with 10 spatially dependent observations is equivalent to 6 pseudo-independent observations.  The growth of the equivalent independent observations is what justifies, asymptotically, the consistency of the Laplace approximations.  Meaning, if the correlation structure of $\Sigma(\theta)$ is strong, then increasing the number of observations may only have minimal impact on the validity of the Laplace approximations. 

In this manuscript we will re-examine some of the shortfalls of using Laplace approximations for inference of spatial or spatio-temporal diffusion parameters.  For a class of models which we will refer to as the self-exciting Poisson CAR models we will show how the assumptions for the first order Laplace approximations of techniques such as INLA may not hold over the entire parameter space.  We will demonstrate how, in this case, higher order approximations of \cite{shun1995laplace} and \cite{evangelou2011estimation} offer more accurate inference and offer greater consistency in parameter estimation and show how the results are comparable to a fully Bayesian inference using rStan of \cite{gelman2015stan}.

\section{Model}

In this manuscript we write $Z(s_i,t)$ for observed count data on a spatial temporal lattice where $s_i \in \{s_1,s_2,...,s_{n_d}\}$ indexes space and $t\in\{1,2,...T\}$ indexes time.  Defining $\boldsymbol{Z_t} = (Z(s_1,t),Z(s_2,t),...,Z(s_{n_d,t}))^T$, the model we consider is

\begin{align}
& Z(s_i,t) \sim \mbox{Pois}(\lambda(s_i,t)) \label{eq:timeseries2} \\
& E[Z(s_i,t)]=\lambda(s_i,t)\\
& \boldsymbol{\lambda_t} = \exp(\boldsymbol{Y_t})+\eta \boldsymbol{Z_{t-1}}\\
& \boldsymbol{Y_t} \sim \mbox{Gau} (\boldsymbol{\alpha_t},(I_{{n_d},{n_d}}-\boldsymbol{C})^{-1}\boldsymbol{M}).
\end{align}

As above, we define $\boldsymbol{C}$ to be the spatial proximity matrix with entry $(i,j)=\zeta$ if the spatial locations, $s_i,s_j$ are neighbors and $0$ otherwise.  $M$ is a diagonal matrix of dimension $n_d \times n_d$ with diagonal entries $\tau^2$.  In order to ensure positive definiteness of the Gaussian covariance matrix we must have $\zeta \in (\psi_{(1)},\psi_{(n)})$ where $\psi_{(k)}$ is the $k$th largest eigenvalue of $\boldsymbol{C}$.   

Data level dependence, or what is commonly referred to as self-excitation, is present in the model through the addition of the $\eta \boldsymbol{Z_{t-1}}$ term to the linear predictor of $\boldsymbol{\lambda}$.  The expected number of events at space-time location $(s_i,t)$ then is a summation of the expected events due to an underlying, latent CAR process, as well as events due to repeat or copy-cat actors.  A sufficient condition to ensure a valid joint density exists is $\eta \in (0,1)$.  

The data model for $Z(s_i,t)$, when conditioned on $Z(s_i,t-1)$ and $Y(s_i,t)$, is then Poisson.  In other words, the density of $Z(s_i,t)$ depends on the previously observed $Z(s_i,t-1)$ and a latent, unobserved $Y(s_i,t)$.

This is similar to an AR(1) version of the Poisson Autoregression model of \cite{fokianos2009poisson}, however with the added complication of independent log-normal errors.  This is also a spatial version of the discrete Hawkes-Cox model of \cite{mohler2013modeling} only allowing a time lag of 1.

The latent process model, $Y(s_i,t)$, is a Conditional Auto-Regressive or CAR model given in \cite{cressie2015statistics} and has joint distribution $\boldsymbol{Y_t}\sim \mbox{Gau}(\boldsymbol{\alpha_t},(I_{{n_d},{n_d}}-\boldsymbol{C})^{-1}\boldsymbol{M})$. Statistically this model is interesting as it is both hierarchical and conditionally specified at the data level, not at the process level.  

As well as being statistically interesting this model also arises naturally when the expected count at space-time location $(s_i,t)$ is equal to the expected count due to a spatial latent process, $\exp(Y(s_i,t))$ and the expected count due to self-excitation, $\eta Z(s_i,t-1)$. This can occur, for example in the modeling of violence in a region.  The latent (unobserved) tension in the region may be solely due to geography or demographics observed at a given space and time.  This may be expressed as a function of large-scale variation, $\boldsymbol{\alpha}$ and small scale variation which is captured in the CAR component of the model.  The critical assumption is that the small scale variation only exists in space.  The second cause of violence in a space-time region may be attributed to the "broken windows" effect, or the propensity of violent action to be repeated in, or near, the same geographical region.  That is, once a violent action occurs, there is some probability that that action will generate copy cats.  As a consequence of the model, if we know $\exp(Y(s_i,t))$ and $\eta$, then the expected number of violent events that arise from model\eqref{eq:timeseries2} can be seen as the sum of the expected number of events due to the latent process and the expected number of events due to copy cat actors.

The likelihood associated with this model is given in \eqref{eq:FullLikelihood}.
\begin{equation}
\small L(\eta,\alpha,\zeta,\tau^2|\boldsymbol{Z}) \propto \int_{\boldsymbol{\Omega}_y} \prod_{i=1}^{n}\prod_{t=1}^{T} \exp(-\eta Z(s_i,t-1)-\exp(Y(s_i,t)))\left(\eta Z(s_i,t-1)+\exp(Y(s_i,t))\right)^{Z(s_i,t)} d\mu_{\boldsymbol{Y}}\label{eq:FullLikelihood}.
\end{equation}

 Due to the temporal independence of $\boldsymbol{Y}$, we can simplify this to
 
 \begin{equation}
 \small L(\eta,\alpha,\zeta,\tau^2|\boldsymbol{Z}) \propto \prod_{t=1}^{T}\int_{\boldsymbol{\Omega}_{y_t}} \prod_{i=1}^{n} \exp(-\eta Z(s_i,t-1)-\exp(Y(s_i,t)))\left(\eta Z(s_i,t-1)+\exp(Y(s_i,t))\right)^{Z(s_i,t)} d\mu_{\boldsymbol{Y_t}}\label{eq:FullLikelihood2}.
 \end{equation}

However, practically, this likelihood cannot be directly maximized due to the intractable integral that is taken with respect to the multivariate Gaussian density associated with $\boldsymbol{Y}$.  If the likelihood could be computed, asymptotic normality of the maximum likelihood estimates could be shown along the lines of \cite{fokianos2009poisson}.  As the log-Gaussian term has support on $(0,\infty)$, many of the standard difficulties of similar models are avoided.  For more on the difficulties of the asymptotics of similar univariate models see Chapter 4 of \cite{davis2016handbook}. Critically in \eqref{eq:timeseries2} we must have $\eta \in (0,1)$, ensuring that the temporal dependence dies off at a geometric rate.

Bayesian Monte Carlo Markov Chain (MCMC) methods also are extremely challenging in this set-up as MCMC techniques will generally either involve integrating \eqref{eq:FullLikelihood} or sampling from the latent states.  A similar model was analyzed in \cite{mohler2013modeling} where inference was conducted using Metropolis Adjusted Langevin Algorithm (MALA).  The challenge in using MCMC techniques including MALA is that the dimension of $\Sigma(\theta) \equiv (I_{n_d,n_d}-\boldsymbol{C})^{-1}\boldsymbol{M}$ is potentially quite large.  Any sampling of $Y$ will require thousands of evaluations of the determinant of this matrix as well evaluations of the log-likelihood.  As we will describe in Section 5 this can be sped up through precomputing eigenvalues of $\boldsymbol{C}$ but even with this, it remains potentially painfully slow and unfeasible in the model building phase of analysis.

\section{Laplace Approximation}

An approximation method similar to Integrated Nested Laplace Approximation (INLA) was used to fit a Self-Exciting Poisson SAR model in \cite{2017arXiv170308429C}.  This inferential technique was first recommended in \cite{tierney1986accurate}.  Generically, we let $\pi(.)$ represent a density function and $\pi(.|.)$ represent a conditional density function.  Now, we can approximate $\pi(\theta|Z)$ where $Z$ is the observed data, $Y$ is a latent random variable, and $\theta$ is the set of parameters that inference by using the relationship

\begin{align}
\pi(\theta|Z)\propto \frac{ \pi(Z,Y,\theta)}{\pi_G(Y|Z,\theta)}\bigg\rvert_{Y=Y^*(\theta)},
\end{align}
where $\pi_G(Y|Z,\theta)$ is the Gaussian approximation to the density $\pi(Y|Z,\theta)$.  Both the numerator and the denominator are then evaluated at the mode of $Y$ for a given $\theta$, denoted as $Y^*(\theta)$.  The benefit of this, when applied to \eqref{eq:FullLikelihood} is that it is essentially an integration free method of marginalizing over $Y$.  For \eqref{eq:timeseries2}, this becomes

\begin{align}
\tilde{\pi}(\eta,\zeta,\tau^2,\alpha|\boldsymbol{Z})\propto \frac{\pi(\boldsymbol{Z}|\eta,\boldsymbol{Y})\pi(\boldsymbol{Y}|\alpha,\zeta,\tau^2)\pi(\zeta)\pi(\alpha)\pi(\tau^2)\pi(\zeta)}{\pi_G(\boldsymbol{Y}|\alpha,\eta,\zeta,\tau^2,\boldsymbol{Z})} \label{eq:INLA},
\end{align}

 where $\tilde{\pi}(\eta,\zeta,\tau^2,\alpha|\boldsymbol{Z})$ is an approximation to the marginal posterior density of $\eta,\zeta,\tau^2,\alpha$, and $\pi_G(\boldsymbol{Y}|\alpha,\eta,\zeta,\tau^2,\boldsymbol{Z})$ is a Gaussian approximation to the joint density of the latent state $\boldsymbol{Y}$. 

The Gaussian approximation given in the denominator of \eqref{eq:INLA} is based off of a Taylor series approximation to the log-density of $\pi(\boldsymbol{Z}|\boldsymbol{Y},\eta)$.  That is, $\pi_G(\boldsymbol{Y}|\alpha,\eta,\zeta,\tau^2,\boldsymbol{Z})=\pi(\boldsymbol{Z}|\boldsymbol{Y},\eta)\pi(\boldsymbol{Z}|\eta,\boldsymbol{Y})$.  Specifically, we can write 
\begin{align}
\pi_G(\boldsymbol{Y}|\eta,\zeta,\tau^2,\boldsymbol{Z}) =& (2 \pi)^{n/2} \det(\Sigma(\theta))^{1/2} \exp(-\frac{1}{2}\boldsymbol{Y}^t \Sigma^{-1}(\theta)\boldsymbol{Y}+\sum_{s_i,t} f(\mu(s_i,t))(Y(s_i,t))+ \nonumber 
\\ \label{eq:gausapprox}& 1/2 k (\mu(s_i,t))(Y(s_i,t))^2)\\ 
\mbox{where in above}\nonumber\\
f(\mu(s_i,t))  =& \frac{Z(s_i,t)\exp(\mu(s_i,t))}{\exp(\mu(s_i,t))+\eta Z(s_i,t-1)}-\exp(\mu(s_i,t)) - \nonumber \\       & \mu(s_i,t)\left(\frac{Z(s_i,t)\exp(\mu(s_i,t))}{\exp(\mu(s_i,t))+\eta Z(s_i,t)}-\frac{\exp(2 \mu(s_i,t))Z(s_i,t)}{\left(\exp(\mu(s_i,t))+\eta Z(s_i,t-1)\right)^2}-\exp(\mu(s_i,t))\right)
\\ \nonumber
k(\mu(s_i,t)) =& -\frac{Z(s_i,t)\exp(\mu(s_i,t))}{\exp(\mu(s_i,t))+\eta Z(s_i,t)}+\frac{\exp(2 \mu(s_i,t))Z(s_i,t)}{\left(\exp(\mu(s_i,t))+\eta Z(s_i,t-1)\right)^2}+\exp(\mu(s_i,t)). \\ 
\end{align}
The expressions $f(.)$ and $k(.)$ given are derived from expanding the log-density of $Z$ as a function of $Y$ about an initial guess for the mode. \eqref{eq:gausapprox} is then maximized as a function of $\boldsymbol{Y}$ and then evaluated at that value.  The computational burden comes in conducting the maximization, however the sparsity of $\Sigma^{-1}(\theta)$ makes this easier, an explicit formula is given in \cite{rue2009approximate}.  

 When \eqref{eq:gausapprox} is evaluated at the posterior mode, it becomes $2 \pi^{n/2} \det (W+\Sigma^{-1}(\theta))^{\frac{1}{2}}$ where $W$ is a diagonal matrix of the same dimension as $\Sigma(\theta)$ where each diagonal entry is $k(\mu(s_i,t))$.  The numerator of \eqref{eq:INLA} is then evaluated at $\mu(s_i,t)$.  Therefore, the problem is simply a computation once the posterior mode of the denominator is found.

Inference is then carried out by fixing values of $\eta,\zeta,\tau^2,\alpha$, then finding the values of $\boldsymbol{Y}$ that maximize the Gaussian approximation.  Then, for those fixed parameter values, we obtain an estimate of the posterior probability.  The parameter space for $\eta,\zeta,\tau^2,\alpha$ can be efficiently explored to map out the marginal likelihood surface for that set of parameters.  \cite{rue2009approximate} discuss efficient methods for exploring the parameter space.

From $\tilde{\pi}(\eta,\zeta,\tau^2,\alpha|\boldsymbol{Z})$ and $\pi_G(\boldsymbol{Y}|\alpha,\eta,\zeta,\tau^2,\boldsymbol{Z})$ we can then estimate the marginal posterior density $\pi(Y|Z)$ by calculating $\pi(Y|Z) \approx \sum\tilde{\pi}(\eta,\zeta,\tau^2,\alpha|\boldsymbol{Z})\pi_G(\boldsymbol{Y}|\alpha,\eta,\zeta,\tau^2,\boldsymbol{Z})$ where the summation is over all values of $\theta$ with sufficiently high posterior probability.  If inferential concern is on the density of the latent state, we can subsequently improve $\pi_G(\boldsymbol{Y}|\alpha,\eta,\zeta,\tau^2,\boldsymbol{Z})$ by using a skew-Normal approximation based off of a higher order Taylor series expansion as given in \cite{rue2009approximate}.

While \eqref{eq:INLA} is a method to conduct Bayesian analysis, in the absence of $\pi(\zeta)$, $\pi(\alpha)$, $\pi(\tau^2)$, the maximization in \eqref{eq:INLA} is also an estimate of the maximization of the likelihood for $\eta$, $\zeta$,$\alpha$ and $\tau^2$ marginalized over $\boldsymbol{Y}$.  Clearly the Gaussian approximation, and hence the Laplace approximation, is asymptotically valid if the Taylor series of $Z$ has a vanishing third and higher derivatives.  Otherwise, the practitioner must rely on the assumption that the higher order terms are negligible. 

\subsection{Issues with Laplace Approximation for Spatio-Temporal Data}
There are two primary concerns with using this technique.  The concerns are somewhat addressed in \cite{rue2009approximate}, but we will make them clear here.  The first concern is unavoidable in any parametric modeling of spatio-temporal data.  To see this issue, it is instructive to consider spatial sampling with temporal replication where there is no temporal dependence.  If we only consider $Z(s_i)$ with $s_i\in \{s_1,s_2,...,s_n\}$ and say we sample this $T$ times, then we have replication of any spatial patterns to conduct inference from.  Without replication, we have to hope that our spatial domain is large enough to create internal replication, that is, that the dependency in the data decays at a sufficient rate.  This same issue exists in spatio-temporal data.  Now, we have data that has dependence in both space and time and we inevitably only have a single realization of the data.  Therefore, our space-time observation must be large enough to break both the space dependence and the time dependence.  Essentially, this means that our, unobservable, space-time clusters must be small.

This is an issue with using Laplace approximations as the inferential results are asymptotically justified through the growth of independent samples.  The approximation error of \cite{tierney1986accurate} is $\mathcal{O}(n^{-3/2})$, however the meaning of '$n$' for spatio-temporal models is not well-defined.   The asymptotics are clearly justifiable if both the size of the grid and the number of observations per node increases, but the $n$ that needs to grow is the number of independent space-time observations.

One method of examining whether this has occurred is to look at the effective number of parameters as defined in \cite{spiegelhalter1998bayesian}.  If the data is completely independent, then $n$ is indeed the number of samples.  In this case, the effective number of parameters is the number of large scale parameters in the model.  If we examine the ratio of observations to the effective number of parameters we will get an estimate of the number of observations available to estimate each of the effective number of parameters.  If, for example, the effective number of parameters is close to $n$, then the ratio of observations to effective number of parameters will be extremely small indicating that we lack sufficient observations to conduct meaningful analysis.

The above concern really applies for any analysis of space-time data when we directly work with the full log-likelihood.  In order to conduct meaningful inference we need to have replication or pseudo-replication of our data.  The second issue is more specific to Laplace approximations and appears to be more prevalent in count data.  That is, there is a bias in the approximation due to the truncation of the Taylor series that underlies the Gaussian approximation in the denominator of \eqref{eq:INLA}.  This appears to first have been demonstrated in \cite{joe2008accuracy} where clustered (temporal) count data was analyzed assuming a Poisson-log Gaussian mixture where the log Gaussian was assumed to have an AR(1) structure.  In \cite{joe2008accuracy}, the AR(1) parameter was consistently shown to be biased low and, assuming zero intercept, the variance was biased high.  \cite{carroll2015comparing} also demonstrated bias in the estimation of the Intrinsic Conditional Auto-Regressive (ICAR) parameter when using the INLA software.  \cite{rue2009approximate} recognize the bias in Laplace approximations, but state that it tends to be negligible in practice and only appear in pathological cases.  However, as we will demonstrate, issues with truncation of the Taylor series approximation underlying the Laplace approximation are a major concern for self-exciting Poisson models like \eqref{eq:timeseries2} for parameter values that arise in practice.

\section{Extended Laplace Approximation}

The primary issue in \eqref{eq:INLA} when applied to \eqref{eq:timeseries2} is that we are essentially conducting a Laplace approximation to an integral of the form 
\begin{equation}
M=\int_{\mathbb{R}^{n_d \times T}} \exp \left(-g(Y|Z,\eta,\zeta,\tau^2,\alpha)\right) dY \label{eq:Integral},
\end{equation}

where 
\begin{equation}
\scriptstyle g(Y|.)= \frac{1}{2}\boldsymbol{Y}^T \Sigma^{-1}(\theta)\boldsymbol{Y}-\left(\sum_{i=1}^{n_d} \sum_{t=1}^T -\eta Z(s_i,t-1)-\exp(Y(s_i,t))+Z(s_i,t)\log\left[\eta Z(s_i,t-1)+\exp(Y(s_i,t))\right]\right) \label{eq:g}. 
\end{equation}

Clearly the size of $g(.)$ matches the dimension of the integration.  As demonstrated in \cite{shun1995laplace}, this results in a necessarily biased approximation to the integral where the bias is on the order of $O(1)$.  

In order to correct these issues, \cite{shun1995laplace} and \cite{evangelou2011estimation} conduct an expansion of $\log (M)$ that is correct even when the dimension of the integral in \eqref{eq:Integral} is equal to the sample size.  The asymptotic behavior then will be appropriate as $T \to \infty$ due to the geometric decay in time induced by $\eta \in (0,1)$.

We will use the notation of \cite{evangelou2011estimation} letting $g_i(Y)= \frac{\partial g(Y)}{\partial Y(s_i,t)}$ and $g_{i,j}(Y)=\frac{\partial^2 g(Y)}{\partial Y(s_i,t) \partial(Y(s_j,t))}$.  We will also let $g_{\boldsymbol{Y}}$ be the gradient of $g$ and $g_{\boldsymbol{YY}}$ be the Hessian and $g^{i,j}$ be the $(s_i,s_j)$ element of the inverse of the Hessian matrix.  

In order to correct for the bias we apply the expansion given as (9) in \cite{shun1995laplace} and (21) in \cite{evangelou2011estimation}.  The correction requires the derivation of the third, fourth and sixth derivatives of $g$, 

\begin{align}
g_{iii}=  &-\left[\exp(Y(s_i,t))\left(\frac{\eta Z(s_i,t-1)}{\lambda(Y(s_i,t))-1}\right)-3\exp(2 Y(s_i,t))\left(\frac{\eta Z(s_i,t-1)}{\lambda(Y(s_i,t))^2}\right)+\right.\nonumber\\
& \left. 2\exp(3Y(s_i,t)))\left(\frac{\eta Z(s_i,t-1)}{\lambda(Y(s_i,t))^3}\right)\right] \label{eq:thirds}
\end{align}

\begin{align}
g_{iiii}=  &-\left[\exp(Y(s_i,t))\left(\frac{\eta Z(s_i,t-1)}{\lambda(Y(s_i,t))-1}\right)-7\exp(2 Y(s_i,t))\left(\frac{\eta Z(s_i,t-1)}{\lambda(Y(s_i,t))^2}\right)+\right.\nonumber\\
& \left. 12\exp(3Y(s_i,t)))\left(\frac{\eta Z(s_i,t-1)}{\lambda(Y(s_i,t))^3})\right)-6\exp(4Y(s_i,t))\left(\frac{\eta Z(s_i,t-1)}{\lambda(Y(s_i,t))^4}\right)\right] \label{eq:fourths}
\end{align}

\begin{align}
	g_{vi}=  &-\left[\exp(Y(s_i,t))\left(\frac{\eta Z(s_i,t-1)}{\lambda(Y(s_i,t))-1}\right)-31\exp(2 Y(s_i,t))\left(\frac{\eta Z(s_i,t-1)}{\lambda(Y(s_i,t))^2}\right)+\right. \nonumber \\
	& 180\exp(3Y(s_i,t)))\left(\frac{\eta Z(s_i,t-1)}{\lambda(Y(s_i,t))^3})\right)-438\exp(4Y(s_i,t))\left(\frac{\eta Z(s_i,t-1)}{\lambda(Y(s_i,t))^4}\right)]+\nonumber\\
	& \left. 408\exp(5Y(s_i,t))\left(\frac{\eta Z(s_i,t-1)}{\lambda(Y(s_i,t))^5}\right)-120\exp(6Y(s_i,t))\left(\frac{\eta Z(s_i,t-1)}{\lambda(Y(s_i,t))^6}\right)\right] \label{eq:sixths}
\end{align}

where $\lambda(Y(s_i,t))=\exp(Y(s_i,t))+\eta Z(s_i,t-1)$ in \eqref{eq:fourths} and \eqref{eq:thirds}.  The final pieces needed are $g^{i,i}$ and $g^{i,j}$ both of which can be found in the appropriate entry upon inverting $\Sigma^{-1}(\theta)+W$ where $W$ is the same as defined in \eqref{eq:gausapprox}, which is the equivalent of $g_{i,i}$.  The evaluation of $\log M$ is then

\begin{align}
\log M & \propto -\frac{1}{2}|\Sigma(\theta)|-\hat{g}-\frac{1}{2}|\hat{g}_{\boldsymbol{YY}}|-\sum_{t}\sum_{i}\frac{1}{8}\hat{g}_{iiii}-\sum_{t}\sum_{i}\frac{1}{48}\hat{g}_{vi} + \nonumber \\
& \frac{1}{72}\sum_{t}\sum_{i,j\leq i}\hat{g}_{iii}\hat{g}_{jjjj}\left(6 \left(\hat{g}^{ij}\right)^3+9 \hat{g}^{ii}\hat{g}^{jj}\hat{g}^{ij}\right) \label{eq:shunExpansion}
\end{align}

In \eqref{eq:shunExpansion} we denote $\hat{g}$ as the evaluation of the $g$ function at $Y(s_i,t)=\mu(s_i,t)$ where $\mu(s_i,t)$ is the point that maximizes the Gaussian approximation to $Y$ in the denominator of \eqref{eq:INLA}.  
 
The evaluation of \eqref{eq:shunExpansion} at this point brings the error from $\mathcal{O}(n^{-1})$ in the Laplace approximations to the marginals, to approximately $\mathcal{O}(n^{-3})$ when the higher order terms are included. While again this $n$ is ill-defined, critically it is the same for both the original and the extended Laplacian, meaning if there is insufficient data to accurately estimate the marginals under \eqref{eq:INLA}, the further expansion may be an improvement.

An alternative would be to employ the derivations in \cite{raudenbush2000maximum} which involve an expansion of $M$ vice $\log M$.  However, as mentioned in \cite{shun1995laplace} and empirically demonstrated in Tables 2 and 3 of that manuscript, this correction has relative error $O(1)$ whereas the correction in \eqref{eq:shunExpansion} has relative error $o(1)$.

The performance of the extended LA method has previously been conducted in a likelihood setting.  The higher order expansion has been shown to provide comparable errors as Gauss-Hermite quadrature with 20 quadrature points (\cite{raudenbush2000maximum}) and Monte Carlo maximum likelihood (\cite{evangelou2011estimation}).   

\subsection{General Algorithm For Conducting Bayesian Inference Using Higher Order Laplace Approximation}

Here we will outline the general algorithm for using \eqref{eq:shunExpansion} to conduct an approximate Bayesian inference for the set of parameters, $\theta=\left(\alpha,\eta,\zeta,\tau^2\right)$.  The first task is finding the mode of $\pi(\theta|\boldsymbol{Z})$.  First we fix a value of $\theta$ and for that value of $\theta$ find the value of $\boldsymbol{Y}^*=\boldsymbol{\mu}^*$ that maximizes \eqref{eq:gausapprox}.  This is accomplished through repeatedly solving $\left(\Sigma^{-1}(\theta)+\text{diag }k(\mu^*(s_i,t))\right)\boldsymbol{\mu}^*=f(\boldsymbol{\mu}^*)$ where $f(\boldsymbol{\mu^*})$ is the vector of evaluations of $f$ given in \eqref{eq:gausapprox}.  The sparsity of $\Sigma^{-1}(\theta)+\text{diag }(k(\mu^*(s_i,t)))$ makes this task extremely fast.

This value, $\boldsymbol{Y}^*$, is then used to evaluate \eqref{eq:thirds}, \eqref{eq:fourths}, and \eqref{eq:sixths}, giving an approximation to the Log-likelihood given in \eqref{eq:shunExpansion}.  As a point of comparison, on a $10 \times 10$ lattice wrapped on a Torus with 100 observations, finding $\boldsymbol{Y}^*$ and computing \eqref{eq:shunExpansion} take approximately 1-1.5 seconds.  Using finite differences, the Hessian at that point can then be approximated.  This takes an additional 32 evaluations if one covariate is in the model.  A Newton-Raphson algorithm can then be used to find the mode of $\tilde{\pi}(\theta|\boldsymbol{Z})$.  In the majority of problems considered, this took us approximately 4-5 steps.  Finding the mode, again for the 10000 size data set described above this, generally, takes about 10-30 minutes.

At the mode, the posterior parameter space can then be efficiently explored using methods outlined in \cite{rue2009approximate}.  Credible intervals for individual elements of $\theta$ can be found either through assuming posterior normality and using the Hessian at the posterior mode or through the method outlined in \cite{ferkingstad2015improving}.  

In summary, the primary advantage of using Laplace based techniques is computational speed.  A single computation of the log-likelihood for a 10 $\times$ 10 neighborhood structure with $T=100$ with $\boldsymbol{\alpha}$ as intercept only takes approximately 1 second with the primary computational cost being incurred in finding the mode of the Gaussian approximation to the denominator of \eqref{eq:INLA}.  In using the extended Laplace approximation method in \eqref{eq:shunExpansion} there is an additional cost of about .5 of a second per evaluation.  As a full exploration of the parameter space may take 600 to 1000 evaluations, the total cost incurred through using the expansion is about 5 to 6 minutes.
\section{Fully Bayesian Approach}

While the size of $\Sigma(\theta)$ makes MCMC techniques challenging, some properties of the model make it feasible to use a flexible modeling language such as Stan to perform inference.  To do this, we follow closely the development given in \cite{joseph}.  First, note that we are trying to find
\begin{align}
\pi(\theta | \boldsymbol{Z})\propto \prod_{s_i,t} \pi(Z(s_i,t)|Y(s_i,t),Z(s_i,t-1),\eta) \pi(Y(s_i,t)|\boldsymbol{\alpha},\tau,\zeta)\pi(\eta)\pi(\boldsymbol{\alpha})\pi(\tau)\pi(\zeta)
\end{align}

In the above, we are required to both sample from and calculate the density of the latent state, $\boldsymbol{Y}$ which requires evaluations of

\begin{align}
\log(\pi(Y(s_i,t)|\boldsymbol{\alpha},\tau,\zeta)) \propto \frac{-t \times n_d}{2}\log(\tau^2) + \frac{1}{2} \log | \Sigma_f^{-1}(\theta)| - \frac{1}{2}(Y-\alpha)^T\Sigma_f^{-1}(\theta)(Y-\alpha) \label{eq:log Y}
\end{align}

To speed up computations, we note that the greatest computational cost in the sampling is the calculation of the determinant of the potentially very large matrix, $\Sigma_f^{-1}(\theta)$.  However, the specific structure for $\Sigma_f^{-1}(\theta)$ allows us to follow \cite{jin2005generalized}.  First we note that $\log | \Sigma_f^{-1}(\theta)|  = T \log | \Sigma^{-1}(\theta)|$ and $\log|\Sigma^{-1}(\theta)|=\frac{n_d}{\log\tau}+\log|I_{n_d,n_d}-\zeta N|$ where $N$ is the neighborhood or adjacency matrix.  Therefore, we can let $V \Lambda V^T$ be the spectral decomposition of $N$ and then $|I_{n_d,n_d}-\zeta N|=|V| |I_{n_d,n_d}-\zeta \Lambda| |V^T|=\prod_{j=1}^{n_d}\left(1-\zeta \lambda_j\right)$ where $\lambda_j$ are the eigenvalues of the neighborhood matrix.  

The greatest advantage of this approach is that the eigenvalues are irrelevant of any parameters, therefore they can be computed ahead of time.  This means that we never need to deal with matrices of the size of  $\Sigma_f(\theta)$.

However, even using state of the art MCMC software such as Stan and precomputing all eigenvalues, MCMC still remains slow.  For example, if $n_d=100$ and $T=100$, a single MCMC chain of length 5000 took 3.5 hours to converge.  In this example, the chain hadn't converged after 1000 iterations but exhibited no signs of non-convergence after 5000.  In comparison, the Laplace approximation method of section 3, under the same set up, takes less than 10 minutes to find the find the parameters that maximize \eqref{eq:INLA} and then another 15-20 minutes to evaluate the posterior parameter space.  The expanded Laplace approximation incurs an additional cost of about .5 of a second per evaluation and under the above conditions would add about 5 to 6 minutes of computations.

\section{Simulation Study}

In order to compare the Laplace approximation, with the higher order Laplace approximation and the MCMC inferential methodology, we simulated data from model \eqref{eq:timeseries2} on a $10 \times 10$ grid wrapped on a torus to reduce edge effects using a rook neighborhood structure.  We further set $t \in \{1,2,...,100\}$.  The choice of these values was made to replicate potential real world situations.  For example, counts aggravated over counties in a state or aggregated over neighborhoods in a major metropolitan area often have approximately 100 locations.  For instance, there are 99 counties in Iowa, there are 96 named neighborhoods in Chicago, and there are 120 districts in Iraq.  $T=100$ would correspond to approximately two years of data observed weekly.

Next, we simulated from all 32 combinations of $\eta \in \{0,.1,.2,.3,.4,.5,.6,.7\}$ and $\tau^2 \in \{.4,.6,.8,1\}$.  For each choice of $\eta$ and $\tau^2$ we next set $\zeta=.245$ in order to generate significant spatial correlation as the spatial correlation.  While we could have considered other choices of $\zeta$ note that the spatial correlation between two observations at the same point in time is
\begin{equation}
\mbox{Corr}(Z(s_i,t)Z(s_j,t))  = \frac{\left(\exp(\Sigma_{i,i}+\Sigma_{i,j}) -\exp(\Sigma_{i,i})\right)}{\left(\exp(2\Sigma_{i,i}) -\exp(\Sigma_{i,i}) + \frac{\exp(-\alpha)}{1-\eta}\exp(\frac{\Sigma_{i,i}}{2}\right)}\label{eq:spatCorr},
\end{equation}
where $\Sigma_{i,j}$ is the $(i,j)$th entry in the covariance matrix, $\Sigma(\theta)$.  In order to have significant correlation in \eqref{eq:spatCorr} $\zeta$ needs to be near the edge of the parameter space.  The spatial correlation reflects a well known problem for CAR models and is presented in depth in \cite{wall2004close}.  We further fixed $\alpha(s_i,t)=0,\forall s_i,t$. 

 For each of the 32 combinations of parameters we found the values of $\hat{\tau^2}$, $\hat{\eta}$ and $\hat{\zeta}$ that maximized \eqref{eq:INLA} and \eqref{eq:shunExpansion}.  In all cases, estimates of $\eta$, $\alpha$, and $\zeta$ using Laplace approximation and expanded LA were generally unbiased.  The difficulty lies in estimating the conditional variance, $\tau^2$.  For even small values of $\eta$ it will be shown that the Laplace expansion used in \eqref{eq:INLA} yields substantial bias.

We define substantial bias as a relative bias that is greater than 15\% of the value of the parameter it is estimating.  For example, if $\tau^2=1$, a substantial bias would exist if the estimation procedure obtained a value greater than 1.15 or less than .85.  We further make the assumption that, all things being equal, \eqref{eq:INLA} is preferable over \eqref{eq:shunExpansion} due to the simplicity of calculating \eqref{eq:INLA}.  We further assume that both of these techniques are preferable over MCMC techniques as they are considerably quicker to fit.

Three example of the results for three combinations of $\eta$ and $\tau^2$ is given in Table \ref{Simulations} with the results from one simulation from each combination.  We further explored the impact of not including \eqref{eq:sixths} in the computation of \eqref{eq:shunExpansion}.  For the MCMC technique, the full parameter space was explored and then the posterior mean was used as a point estimate.  In all cases, vague proper priors were used for $\eta$, $\tau^2$, $\alpha$ and $\zeta$. 

\begin{table}[h]
	\begin{center}
		\begin{tabular}{ |c|c|c|c| } 
			\hline
		& $\eta=.1$, $\tau^2=.4$&$\eta=.4$, $\tau^2=.6$& $\eta=.7$, $\tau^2=1$\\
			\hline
			Relative Bias in LA(1) & .12 & .2 & .46\\ 
			Time to Fit LA(1) (min.)& 10-15 & 16-20 & 16-20\\
			\hline
			Extended LA Without 6th Order & .03 & .1 & .2\\
			Extended LA With 6th Order& .03 & .05 & .2\\
			Time to Fit Extended LA & 20-30 & 20-30 &  25-35\\
			\hline
			MCMC & .02 & .02 & .06\\
			Time to Fit MCMC & 150-250 & 400-650 & 500-650\\
			\hline
		\end{tabular}
	\end{center}
	\caption{Relative Bias and approximate times to find point estimates.  Note that the MCMC time is for a full exploration of the parameter space.  All time to fit are estimates and in the case of LA(1) and Extended LA they are dependent on initial guess for Newton Raphson algorithm.  In general the fit times between LA(1) and the Extended LA are comparable while MCMC took 2-10 hours depending on the simulation run.}\label{Simulations}
\end{table}

\begin{figure}[!htp]
	\centering
	\includegraphics[width=0.5\linewidth, height=0.3\textheight]{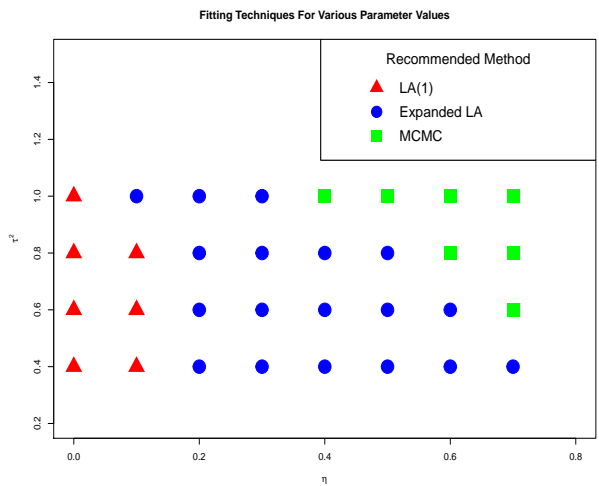}
	\caption{Preferred Methods for inference }\label{fig:Fitting}
\end{figure}

In figure \ref{fig:Fitting}, we display, for all parameter combinations, the preferred method for inference.  As a general algorithm for fitting, we would first attempt the LA(1) approximation.  If the value of $\eta$ or $\tau^2$ is sufficiently high, then we would use the expanded LA method.  Only in cases for extreme $\eta$ and $\tau^2$ would MCMC be necessary.

\begin{figure}[!htp]
	\centering
	\includegraphics[width=0.5\linewidth, height=0.3\textheight]{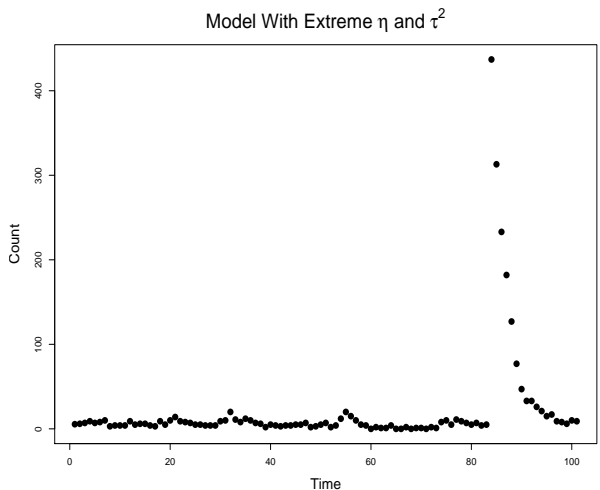}
	\caption{Counts from a simulated location with $\eta=.7$ and $\tau^2=1$}\label{fig:Extreme}
\end{figure}

As depicted in figure \ref{fig:Fitting} for $\eta < .4$ and $\tau^2<1$, the extended Laplace approximation method outlined in Section 1 would offer significant capability to produce correct estimates of parameters.  While this may seem like a strong restriction on the parameter space, values larger than $\eta> .6$ results in extremely peaked and variable data, of which is rarely seen in the cases we envision the self-exciting Poisson CAR model being used.  For example, if we simulate with $\tau^2=1$ and $\eta=.7$, the resulting simulation from a single node is depicted in Figure \ref{fig:Extreme}.  As shown here, these parameter settings would correspond to a situation where there where very low counts followed by a massive spike and slow decay back to low counts.  If the model were to be used to model something like the number of violent crimes in a neighborhood, it would be extremely unlikely that the data would follow this pattern.

\section{Illustrative Example}

In the following section we consider modeling violent crime in the city of Chicago in 2015 using the Self-Exciting Poisson CAR model.  The Self-Exciting Poisson CAR model is appropriate here as there are potentially multiple processes that are giving rise to the violence.  Specifically, some crime may be due to a latent tension at a given location and there may be further violence that is due to copy-cat or retaliatory attacks.  Previous work including \cite{mohler2013modeling} analyzed this data in the absence of spatial correlation and concluded that self-excitement was present.  Our purpose here is not to fully explore the complex nature of how and why violence occurred in Chicago, but rather to demonstrate how the expanded LA could be used by social scientists to quickly explore competing theories within the Self-Exciting Poisson CAR framework allowing the practitioner to capture latent spatial correlation while allowing for the possibility of self-excitation.

The data used for the Chicago crimes is provided via \url{https://data.cityofchicago.org/Public-Safety/Crimes-2001-to-present/ijzp-q8t2}.  We then aggregated all violent crimes both weekly and within specific predefined neighborhoods.  We considered aggravated assault, aggravated battery, and homicides involving weapons as violent crimes.  While there are certainly other violent crimes that could be considered, these crimes in particular seem likely to exhibit self-excitation within a given neighborhood as they potentially spur some form of retaliation.  Similar data was used in both \cite{mohler2013modeling} and \cite{mohler2014marked}. 

While there are no official neighborhoods in Chicago and counts can vary between 77 and 200 named areas, the city of Chicago publishes boundaries at \url{https://data.cityofchicago.org/browse?q=neighborhoods&sortBy=relevance} of 77 distinct neighborhoods.  These are the neighborhoods we used in the analysis and appear to be consistent with historical norms for both locations and naming conventions within the city.  We are not aware of previous statistical studies analyzing crime aggregated to neighborhood levels within the Chicago to compare the choice of neighborhood structure to.  \cite{mohler2013modeling} used data within a specific police beat, which corresponds, approximately, to half or a third of the size of one of the neighborhoods.

\begin{figure}[!htp]
	\centering
	\includegraphics[width=0.5\linewidth, height=0.3\textheight]{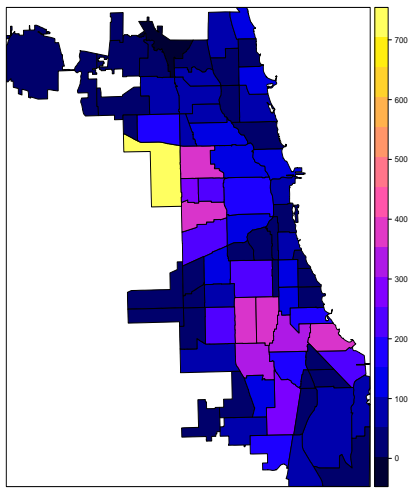}
	\caption{Total count of violent crimes for 2015 aggregated over neighborhood.}\label{fig:SpatialOnly}
\end{figure}

The resulting dataset consists of 9237 violent crimes that occurred in the city over 53 weeks (December 28 2014 - January 2, 2016).  A spatial map depiction of the crimes aggregated over neighborhoods is given in Figure \ref{fig:SpatialOnly}.

As evident in Figure \ref{fig:SpatialOnly}, there appears to be spatial clustering in both the south and the western regions of the city.  Spatial tests such as Moran's I applied to the aggregated data suggest clustering in space and time.  As the data is available on block level we can also treat it as point process data and use Ripley's K which echoes the finding of  clustering in both space and time.

We then fit the data using the model given in Section 1 and in \eqref{eq:timeseries3}:

\begin{align}
& Z(s_i,t) \sim Po(\lambda(s_i,t)) \label{eq:timeseries3} \\
& E[Z(s_i,t)]=\lambda(s_i,t)\\
& \boldsymbol{\lambda_t} = \exp(\boldsymbol{Y_t})+\eta \boldsymbol{Z_{t-1}}\\
& \boldsymbol{Y_t} \sim Gau (\boldsymbol{\alpha_t},(I_{{n_d},{n_d}}-C)^{-1}M)
\end{align}

A well-known phenomenon in criminology, as shown in \cite{anderson1987temperature}, is that higher temperatures are related to higher levels of both violent and non-violent crimes.  To control for this, structure was placed on $\boldsymbol{\alpha_t}$.  Specifically, for location $(s_i,t)$, $\alpha(s_i,t)=\beta_0 + \beta_1 x_1(s_i,t) + \beta_2 x_2(s_i,t)$ where $x_1(s_i,t)$ corresponds to the observed average temperature in neighborhood $s_i$ and time $t$ and $x_2(s_i,t)$ corresponds to the log-population of location $s_i$ at time $t$.  Due to data limitations, we assume that temperature is constant across neighborhoods at time $t$ and population is constant across time at neighborhood $s_i$.  To aid in estimation of covariates, we centered and scaled the temperatures. We used census data for each neighborhood from the United States Census Bureau in 2010.  For temperature, we used historic temperatures available from the Weather Underground website at \url{www.wunderground.com}.

Using the higher order Laplace approximation given in \eqref{eq:shunExpansion} we used finite differences to build up estimates of the Hessian matrix allowing us to perform approximate Newton-Raphson maximization for the parameter space.  With 6 covariates, $\theta = (\tau^2,\zeta,\eta,\beta_0,\beta_1,\beta_2)$, this is possible in a relatively short amount of time.  On a Surface Pro 3, the maximization was done using the statistical software R in under 10 minutes.  The observed maximum was found at $\hat{\theta}=(.52,.179,.50,-5.6,.18,.49)$.  Point estimates using each inferential technique is given in Table \ref{Table:Results}. 

The positive value of $\beta_1$ observed here echoes the findings of \cite{anderson1987temperature} that increasing temperatures increase the probability of violence occurring.  Specifically, because of the structure of model \eqref{eq:timeseries3}, 
if, for a given neighborhood, the temperature changes from 50 degrees Fahrenheit to 90 degrees Fahrenheit, the model would suggest that the expected number of violent crimes, due to temperature alone, would increase by a factor of 2, when controlling for self-excitement in the model.  

The interpretation of $\eta$ differs slightly than the large scale parameters in $\boldsymbol{\alpha}$.  A value of .49 that each violent events at time period $t$ raises the expected number of events at time period $t+1$ by .49.  In other words, if there were 10 violent events in week 1 at a given location we would expect there to be 5 events in week 2 that were 'copy-cat' or inspired by the violence in week 1.

Confidence intervals can then be constructed either relying on asymptotics of the MLE, or in a Bayesian construct, through efficiently exploring the parameter space of $\pi(\theta|\boldsymbol{Z})$ through techniques outlined in \cite{rue2009approximate}.  Here we rely on exploring the parameter space and calculating $\pi(\theta|\boldsymbol{Z})$ over a wide range of $\theta$ values.  Marginals can then be constructed either naively or through skewness corrections as outlined in \cite{martins2013bayesian}.  Here, we approximated the Hessian at the posterior mode using finite differences.  The expanded Laplace approximation was then used to evaluate each of the finite differences to approximate the second partial derivatives.  This technique resulted in credible intervals of $\tau^2 \in (.43,.61)$, $\zeta \in (.176,.182)$, $\eta \in (.47,.53)$, $\beta_0 \in (-6.3,-4.9)$, $\beta_1 \in (.09,.27)$, and $\beta_2 \in (.42,.55)$.  Credible intervals for each parameter are given in \ref{Table:Results2}.

Goodness of fit can be assessed through the use of a randomized version of uniform residuals for discrete observations obtained through the probability integral transform as outlined in \cite{brillinger1982maximum}.  If we let $z_{[1]},z_{[2]},...$ be the possible values of $Z(s_i,t)$, we set our observed $z(s_i,t)=z_{[k]}$ be the $k$th observed value, then the residuals are found through $r(s_i,t)\equiv u(s_i,t)$ where $u(s_i,t)\sim iid\text{ }\mbox{Unif}(F(z_{[k-1]}|\theta),F(z_{[k]}|\theta))$ where $F(.|\theta)$ corresponds to the CDF of $Z$ marginalized over the posterior density of $\theta$.  Practically, this is done through simulating the CDF through an empirical density after repeatedly randomly drawing values from $\pi(\theta|Z)$.  These generalized residuals should be approximately uniform.

\begin{figure}[!htp]
	\centering
	\includegraphics[width=0.5\linewidth, height=0.3\textheight]{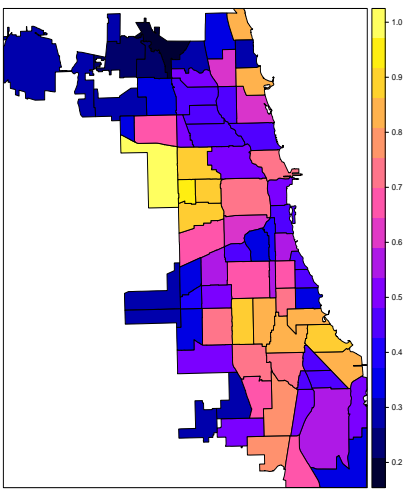}
	\caption{Uniform residuals marginalized over neighborhood.}\label{fig:resids}
\end{figure}

The generalized residuals for this dataset are not uniform when examined against the spatial structure.  If we aggregate the residuals over neighborhood they should be approximately .5 and should have no spatial clustering.  However, if we examine Figure \ref{fig:resids} we see clustering of high residual values in neighborhoods that share similar socio-economic factors.  If we would look at the specific locations of high residuals we would find them in the neighborhoods of Austin, West Garfield Park, and North Lawndale all have high residual values and all have a high percentage of poverty and individuals living on government assistance.  While the socio-economic correlation with violence is not surprising, an analysis of the residuals makes this clear.  This finding suggests that a more detailed investigation of the spatial dimensions of crime in Chicago could be conducted by sociologists who could add relevant spatial structure to $\boldsymbol{\alpha}$ in \eqref{eq:timeseries2}.

To examine the bias in the standard Laplace approximation we next fit to \eqref{eq:timeseries2} using the first-order Laplace approximation method.  Due to the high value of $\eta$ we would expect there to exist a bias in the point estimates.  We again used finite differences to approximate the Hessian and used a Newton-Raphson method to maximize the posterior.  Using this inferential technique the parameters were maximized at $\hat{\theta}=(.38,.180,.50,-5.6,.17,.50)$, again depicted in Table \ref{Table:Results}  Gaussian approximations to the marginals are $\tau^2 \in (.33,.43)$, $\zeta \in (.178,.183)$, $\eta \in (.47,.53)$, $\beta_0 \in (-5.7,-5.4)$, $\beta_1 \in (.11,.23)$, and $\beta_2 \in (.48,.50)$.  As is seen in Table \ref{Table:Results}.  Clearly the largest difference in the point estimation is in $\tau^2$ as the point estimate using LA(1) is over two standard deviations from the estimate using the extended LA method.  Furthermore, 95\% credible intervals for $\tau^2$ do not even overlap, as seen in \ref{Table:Results2}.

Finally, to compare the extended Laplace approximation to an MCMC technique we fit the model approach using the rStan software of \cite{gelman2015stan} using the technique outlined in section 5.  This requires prior specification for all parameters.  In order to be as uninformative as possible, we chose diffuse proper priors.  Specifically, $\pi(\tau)\sim \text{Ca}^{+}(5)$, $\pi(\zeta) \sim \text{Unif}(0,.185)$, $\pi(\eta)\sim \text{Unif}(0,1)$, and $\pi(\beta_0),\pi(\beta_1),\pi(\beta_2) \sim \text{Gau} (0,1000)$.  Where $\text{Ca}^+$ is a half-Cauchy.  The parameter space of $\zeta$ is dictated by the largest eigenvalue of the spatial adjacency neighborhood, in this case the largest eigenvalue is approximately 5.4 constraining $\zeta \leq .185$.  Three chains were run, starting at different locations in the parameter space.  The chains were run for 10000 iterations each.  Stan uses the first half of the iterations for warm-up, resulting in 15000 posterior samples for each parameter.  Convergences was determined through examining the $\hat{R}$ values as well as through visual examination of the trace plots.  Specific for using Stan, the divergence of the chains must be examined, see e.g. \cite{betancort}.  After 10000 iterations there was no evidence that the chains had not converged.  The entire process, using multiple cores to run each chain, took 3 hours.  If parallelizing was not performed, it would take approximately 9 hours to run.

Using MCMC, 95 \% credible intervals were $\tau^2 \in (.42,.59)$, $\zeta \in (.176,.182)$, $\eta \in (.47,.53)$, $\beta_0 \in (-6.3,-5.0)$, $\beta_1 \in (.09,.27)$, and $\beta_2 \in (.42,.56)$.  A comparison of point estimates is given in Table \ref{Table:Results} and a comparison of credible intervals found through MCMC and extended LA is given in Table \ref{Table:Results2}.  As is clearly evident, there is not a significant difference between the extended Laplace technique and MCMC, however the time to fit the model was drastically higher using MCMC.  While LA(1) and the extended LA were fit in similar time, LA(1) appears to underestimate $\tau^2$, which is consistent with what was found during the simulations in Section 6.

\begin{table}[h]
\begin{center}
	\begin{tabular}{ |c|c|c|c|c|c|c| } 
		\hline
		Point Estimates & $\tau^2$ & $\zeta$ & $\eta$ & $\beta_0$ & $\beta_1$ & $\beta_2$\\
		\hline
		LA(1) & .38 & .180 &.50& -5.6 & .17 & .50 \\ 
		Extended LA & .52 & .179 &.50& -5.6 & .18 & .49\\
		MCMC & .50 & .179 & .50 & -5.6 & .18 & .49\\
		\hline
	\end{tabular}
\end{center}
\caption{Point estimates of the parameters from fitting model \eqref{eq:timeseries2} to the Chicago crime data.  As evident, the Expanded LA and MCMC techniques are extremely similar, while LA(1) has a bias for $\tau^2$.}\label{Table:Results}
\end{table}

\begin{table}[h]
	\begin{center}
		\begin{tabular}{ |c|c|c|c|c|c|c| } 
			\hline
			95\% Credible Intervals & $\tau^2$ & $\zeta$ & $\eta$ & $\beta_0$ & $\beta_1$ & $\beta_2$\\
			\hline 
			Extended LA & (.43,.61) & (.176,.182) &(.47,.53)& (-6.3,-4.9) & (.09,.27) & (.42,.55)\\
			MCMC & (.42,.59) & (.176,.182) & (.47,.53) & (-6.3,-5.0) & (.09,.27) & (.42,.56)\\
			\hline
		\end{tabular}
	\end{center}
	\caption{Comparison between 95 \% credible intervals formed using Expanded LA and MCMC.  Note that the 95 \% credible intervals for Expanded LA were donethrough using finite differences to approximate the Hessian and then using a Gaussian approximation to the posterior.}\label{Table:Results2}
\end{table}

\newpage
\section{Discussion}
In this manuscript we demonstrated how extending Laplace approximations to include sixth order derivatives significantly reduces the bias in self-exciting spatio-temporal models.  In general, as long as the marginal variance of the process model $\boldsymbol{Y}$ is less than 1 and the self-excitement parameter is less than .6, the extended Laplace approximations will give estimates that are nearly unbiased, and the bias will reduce as the number of observations per location increases.  We note that \cite{ferkingstad2015improving} also offers a copula based method for potentially correcting the bias, however, this takes the analysis out of the Laplace framework and it is unclear what proceeding along this line does to the asymptotics.  Furthermore, in the example considered in this manuscript, we were not interested in $\boldsymbol{Y}|\boldsymbol{Z}$.  In order to implement the methodology outlined in \cite{ferkingstad2015improving} we would need to calculate the skew-normal approximation to $Y|\theta,Z$ which would add to the computational burdern.

We further showed how a fully Bayesian approach could be considered through exploiting the sparsity of the precision matrix of the spatio-temporal process model.  Even with a fully Bayesian approach being possible, the main benefit of using an extended LA methodology for this model is in computational speed.  While MCMC takes several hours, the entire process for the extended LA took approximately half an hour.  The datasets we considered here were moderately sized for spatio-temporal data, if, however, we used larger datasets we would expect there to be an even larger disparity in fitting time.

The obvious cost of using the extended LA methodology is it requires deriving up to sixth order partial derivatives to compute \eqref{eq:shunExpansion}.  Also, under the methodology outlined in this manuscript, exploration of the parameter space would not be efficient for a higher number of covariates in the model.  However, as demonstrated above, if a Gaussian approximation to the marginals were to be used the parameter space would not have to be fully explored and second order finite differences could be used to fairly quickly approximate the Hessian.

Finally, we demonstrated how this methodology can be applied to analyze crime in Chicago showing how both spatial and temporal covariates can be considered through placing structure on $\boldsymbol{\alpha}$ and in this instance matches the inference using MCMC techniques.  Interestingly, the self-excitement value found in this analysis, $\hat{\eta}=.50$, is similar to what was found in \cite{mohler2013modeling} where in one police beat, 55\% of observed crime was found to be due to repeated actions, or self-excitement.  While that manuscript did not consider exogeneous covariates, our analysis would suggest that the self-excitement was present even when weather and population size were considered.  

While socio-economic factors weren't considered in our analysis, the residuals suggest that researchers with expertise in this area may apply this model with the addition of relevant covariates accounting for these factors.  Significantly, this would allow for inference for these factors controlling for the existence of self-excitement, which appears to be done rarely, if ever, in this field of literature. 
\section{Supplemental Material} 

\begin{description}

\item[Data Sets Used in Illustrative Example] .csv files containing the crime counts aggregated over neighborhoods and weeks, the weather aggregated over neighborhoods and weeks, and the population aggregated over neighborhoods and weeks

\item[chi.graph] Graph file giving the neighborhood structure for the 77 neighborhoods in Chicago

\item[R-code used in RStan] R-code and Stan model to do fully Bayesian method used in Illustrative Example

\end{description}

\clearpage

\bibliographystyle{elsarticle-harv} 
\bibliography{BibFile2}
\end{document}